\documentclass[11pt]{article}
\usepackage{graphicx}
\usepackage{longtable,lscape}
\usepackage{amsfonts}
\usepackage{float}
\usepackage{url}

\newcommand{\captionfonts}{\footnotesize}
\makeatletter  
\long\def\@makecaption#1#2{%
  \vskip\abovecaptionskip
  \sbox\@tempboxa{{\captionfonts #1: #2}}%
  \ifdim \wd\@tempboxa >\hsize
    {\captionfonts #1: #2\par}
  \else
    \hbox to\hsize{\hfil\box\@tempboxa\hfil}%
  \fi
  \vskip\belowcaptionskip}
\makeatother 
\begin{document}
\title{A Quantum-Conceptual Explanation of Violations of \\ Expected Utility in Economics}
\author{Diederik Aerts$^1$, Jan Broekaert$^1$, Marek Czachor$^2$ and Bart D'Hooghe$^1$ \vspace{0.2 cm} \\ 
        \normalsize\itshape
        $^1$ Center Leo Apostel for Interdisciplinary Studies \\
        \normalsize\itshape
        Vrije Universiteit Brussel, 1160 Brussels, 
       Belgium \\
        \normalsize
        E-Mails: \url{diraerts@vub.ac.be,jbroekae@vub.ac.be}, \\  \normalsize \textsf{bdhooghe@vub.ac.be} 
        \vspace{0.1 cm} \\
        \normalsize\itshape
        $^2$ Katedra Fizyki Teoretycznej i Metod Matematycznych \\
        \normalsize\itshape
        Politechnika Gdanska, 80-952 Gdansk, Poland. \\
        \normalsize
        E-Mail: \url{mczachor@pg.gda.pl}
        }
\date{}
\maketitle              
\begin{abstract} 
\noindent The expected utility hypothesis is one of the building blocks of classical economic theory and founded on Savage's Sure-Thing Principle. It has been put forward, e.g. by situations such as the Allais and Ellsberg paradoxes, that real-life situations can violate Savage's Sure-Thing Principle and hence also expected utility. We analyze how this violation is connected to the presence of the `disjunction effect' of decision theory and use our earlier study of this effect in concept theory to put forward an explanation of the violation of Savage's Sure-Thing Principle, namely the presence of `quantum conceptual thought' next to `classical logical thought' within a double layer structure of human thought during the decision process. Quantum conceptual thought can be modeled mathematically by the quantum mechanical formalism, which we illustrate by modeling the Hawaii problem situation, a well-known example of the disjunction effect, and we show how the dynamics in the Hawaii problem situation is generated by the whole conceptual landscape surrounding the decision situation.
\end{abstract}

\section{Introduction}
The mathematical modeling of expected utility in economics starts with the seminal work of John von Neumann and Oskar Morgenstern on economic behavior and games theory \cite{vonneumannmorgenstern1944}. One of the basic principles of the von Neumann-Morgenstern theory is Savage's `Sure-Thing Principle' \cite{savage1954}, which is equivalent to the independence axiom of expected utility theory. Over the years, different modified versions of von Neuman-Morgenstern's original axiomatization of expected utility have emerged, while also its premisses have been criticized by means of specific examples referred to as paradoxes. The Allais paradox \cite{allais1953} and the Ellsberg paradox \cite{ellsberg1961}, for example, point to an inconsistency with the predictions of the expected utility hypothesis, indicating a violation of the independence axiom and the Sure-Thing Principle. `Uncertainty aversion' is generally believed to be at the origin of this violation. 

In recent works, we have analyzed aspects of human thought \cite {aerts2009,aertsdhooghe2009} from the perspective of ongoing investigations on concepts and how they combine, and an approach to use the quantum-mechanical formalism to model such combinations of concepts \cite{aertsbroekaertgabora2010,aertsgabora2005a,aertsgabora2005b,gaboraaerts2002}. In this way, we have shown \cite {aerts2009,aertsdhooghe2009} that two superposed layers can be distinguished in human thought: (i) a layer given form by an underlying classical deterministic process, incorporating essentially logical thought and its indeterministic version modeled by classical probability theory; (ii) a layer given form under the influence of the totality of the surrounding conceptual landscape, where different pieces of this conceptual landscape figure as individual entities rather than as logical combinations of concepts, with measurable quantities such as `typicality', `membership', `representativeness', `similarity', `applicability', `preference' and also `utility' carrying the influences. We have called the process in this second layer `quantum-conceptual thought' \cite
{aerts2009,aertsdhooghe2009}, which is indeterministic in essence, and contains holistic aspects, but is equally well organized as logical thought, although in a very different manner. A substantial part of the `quantum-conceptual thought process' can be modeled by quantum-mechanical probabilistic and mathematical structures.

In the present article, we intend to apply the insights gained into the two-layered structure of human thought, and the role played by the quantum formalism in modeling the decision processes involved in what we have called the quantum-conceptual layer, to analyze the violation of the independence axiom of expected utility theory. More specifically, we will look at the violation of the Sure-Thing Principle connected to what psychologists call the disjunction effect \cite{tverskyshafir1992} and how a quantum-mechanical model can be proposed to describe this type of situation in mathematical terms. We will focus on a well known example of the disjunction effect, the `Hawaii problem' \cite{tverskyshafir1992}, and construct a quantum modeling for the experimental data collected for a test of the Hawaii problem. We also analyze the connection with a disjunction-type effect appearing for the `disjunction of concepts', studied experimentally within concept research \cite{hampton1988}, and illustrate how this analysis indicates that the whole conceptual landscape, hence not only an effect of ambiguity aversion, plays a role in the deviations of classicality. We study experiments with respect to the disjunction effect in the form of a modified Hawaii situation \cite{bagassimacchi2007}, and show how these experiments also show a contextual influence of the whole conceptual landscape, and use the explanations in our work on concept combinations \cite{aerts2009,aertsdhooghe2009,aertsbroekaertgabora2010,aertsgabora2005a,aertsgabora2005b,aerts2007a,aerts2007b} to shed light on the violation of the Sure-Thing Principle.  In \cite{aertsdhooghesozzo2011} we use the approach in the present article to investigate the Ellsberg paradox and its violation of the Sure-Thing Principle. 

\section{The Sure-Thing Principle and the disjunction effect}

Savage introduced the Sure-Thing Principle \cite{savage1954} was inspired by the following story: {\it A businessman contemplates buying a certain piece of property. He considers the outcome of the next presidential election relevant. So, to clarify the matter to himself, he asks whether he would buy if he knew that the Democratic candidate were going to win, and decides that he would. Similarly, he considers whether he would buy if he knew that the Republican candidate were going to win, and again finds that he would. Seeing that he would buy in either event, he decides that he should buy, even though he does not know which event obtains, or will obtain, as we would ordinarily say.} The Sure-Thing Principle is equivalent to the independence axiom of expected utility theory: `independence' here means that if a person is indifferent in their choice between simple lotteries $L_{1}$ and $L_{2}$, they will also be indifferent in choosing between $L_{1}$ mixed with an arbitrary simple lottery $L_{3}$ with probability $p$ and $L_{2}$ mixed with $L_{3}$ with the same probability $p$.

The above situation is similar to what in psychology is called the disjunction effect. A well-known example of this disjunction effect is the so-called Hawaii problem \cite{tverskyshafir1992}, which is about the following two situations.

{\it Disjunctive version}: Imagine that you have just taken a tough qualifying examination. It is the end of the fall quarter, you feel tired and run-down, and you are not sure that you passed the exam. In case you failed you have to take the exam again in a couple of months after the Christmas holidays. You now have an opportunity to buy a very attractive 5-day Christmas vacation package to Hawaii at an exceptionally low price. The special offer expires tomorrow, while the exam grade will not be available until the following day. Would you: $x$ buy the vacation package; $y$ not buy the vacation package; $z$ pay a \$5 non-refundable fee in order to retain the rights to buy the vacation package at the same exceptional price the day after tomorrow after you find out whether or not you passed the exam?

{\it Pass/fail version}: Imagine that you have just taken a tough qualifying examination. It is the end of the fall quarter, you feel tired and run-down, and you find out that you passed the exam (failed the exam. You will have to take it again in a couple of months after the Christmas holidays). You now have an opportunity to buy a very attractive 5-day Christmas vacation package to Hawaii at an exceptionally low price. The special offer expires tomorrow. Would you: $x$ buy the vacation package; $y$ not buy the vacation package: $z$ pay a \$5 non-refundable fee in order to retain the rights to buy the vacation package at the same exceptional price the day after tomorrow.

In the Hawaii problem, more than half of the subjects chose option $x$ (buy the vacation package) if they knew the outcome of the exam (54\% in the pass condition and 57\% in the fail condition), whereas only 32\% did so if they did not know the outcome of the exam.

This Hawaii problem demonstrates clearly a violation of the Sure-Thing Principle. Indeed, subjects prefer option $x$ (to buy the vacation package) when they know that they passed the exam and they also prefer $x$ when they know that they failed the exam, but they refuse $x$ (or prefer $z$) when they don't know whether they passed or failed the exam. The Hawaii problem suggests that the origin of the violation of the Sure-Thing Principle is `uncertainty aversion'. Indeed, subjects prefer to buy the vacation package in both cases where they have certainty about the outcome of the exam, while they refuse to buy the package when they do not yet know whether they passed or failed the exam and hence lack this certainty.

\section{Quantum Modeling of the Hawaii disjunction effect}
The disjunction effect in decision theory is an example of a situation that can be described in the general quantum-modeling scheme that we elaborated in earlier work \cite{aerts2007a,aerts2007b,aerts2009,aertsdhooghe2009,aertsbroekaertgabora2010,aertsgabora2005a,aertsgabora2005b}. In the following, we put forward an explicit example of such a quantum model.
Let us denote by $A$ the conceptual situation in which the subject has passed the exam, and by $B$ the conceptual situation in which the subject has failed the exam. The disjunction of both conceptual situations, denoted `$A$ or $B$', is the conceptual situation where the subject \textit{has passed or failed the exam}. The subject needs to make a decision whether to buy the vacation package (positive outcome) or not to buy it (negative outcome).

In working out our quantum-modeling of the disjunction effect, we use the notion of `state of a concept' in the way it was introduced in \cite{aertsbroekaertgabora2010,aertsgabora2005a,aertsgabora2005b} and represent the above conceptual situations by such states, and hence by a unit vector in a complex Hilbert space, by analogy with how the state of a quantum entity is represented in quantum mechanics. Hence we represent $A$ by a unit vector $|A\rangle $ and $B$ by a unit vector $|B\rangle $ in a complex Hilbert space $\mathcal{H}$, respectively. We take $|A\rangle $ and $ |B\rangle $ orthogonal, hence $\langle A|B\rangle =0$, and describe the disjunction `$A$ or $B$' by means of the normalized superposition state ${1 \over \sqrt{2}}(|A\rangle +|B\rangle )$. The decision to be made is `to buy the vacation package' or `not to buy the vacation package'. This decision is described in our quantum-modeling scheme by means of projection operator $M$ of the Hilbert space ${\cal H}$. The probability for an outcome `yes' (buy the package) in the `pass' situation (state $|A\rangle$) is 0,54, and let us denote this probability by $\mu(A) = 0,54$. The probability for an outcome `yes' (buy the package) in the `fail' situation (state $|B\rangle$) is 0.57, i.e. in our notation $\mu(B) = 0.57$. The probability for an outcome `yes' (buy the package) in the `pass or fail' situation (state ${1 \over \sqrt{2}}(|A\rangle +|B\rangle )$) is 0.32, i.e. in our notation $\mu(A\ {\rm or}\ B) = 0.32$. 

In accordance with the quantum rules we have 
\begin{equation}
\mu (A)=\langle A|M|A\rangle \quad \mu (B)=\langle B|M|B\rangle \quad \mu
(A\ \mathrm{or}\ B)={\frac{1}{2}}(\langle A|+\langle B|)M(|A\rangle
+|B\rangle )  \label{quantprob}
\end{equation}%
Applying the linearity of Hilbert space and taking into account that $%
\langle B|M|A\rangle ^{\ast }=\langle A|M|B\rangle $, we have 
\begin{eqnarray}
\mu (A\ \mathrm{or}\ B) &=&{\frac{1}{2}}(\langle A|M|A\rangle +\langle
A|M|B\rangle +\langle B|M|A\rangle +\langle B|M|B\rangle )
\label{interference} \\
&=&{\frac{\mu (A)+\mu (B)}{2}}+\Re \langle A|M|B\rangle 
\end{eqnarray}%
where $\Re \langle A|M|B\rangle $ is the real part of the complex number $%
\langle A|M|B\rangle $, i.e. the interference term. Its presence allows to
produce a deviation from the average value ${\frac{1}{2}}(\mu (A)+\mu (B))$,
which would be the outcome in the absence of interference. Note that we have
applied two of the key quantum features, namely \textit{superposition}, in
taking ${\frac{1}{\sqrt{2}}}(|A\rangle +|B\rangle )$ to represent `$A$ or $B$%
', and \textit{interference}, as the effect appearing in equation (\ref%
{interference}).

This `quantum model based on superposition and interference' can be realized
in a three-dimensional complex Hilbert space ${\mathbb{C}}^{3}$. For a more
detailed analysis we refer to \cite{aerts2007a,aerts2007b,aerts2009}. Here we focus on the ${\mathbb{C}}^{3}$ realization for the
Hawaii problem. We remark that, for a given $\mu (A)$ and $\mu (B)$, the
situation is always such that $\mu (A)+\mu (B)$ is smaller than or equal to
1 or is greater than 1. Let us now construct explicitly the ${\mathbb{C}}^{3}$ model. In case (i) $\mu (A)+\mu (B)\leq 1$, we put $a=1-\mu
(A),$ $b=1-\mu (B)$ and $\gamma =\pi $, and in case (ii) $1<\mu (A)+\mu (B)$%
, we put $a=\mu (A)$, $b=\mu (B)$ and $\gamma =0$. We choose 
\begin{eqnarray}
|A\rangle &=&(\sqrt{a},0,\sqrt{1-a})  \label{vectorA} \\
|B\rangle &=&e^{i(\beta +\gamma )}(\sqrt{\frac{(1-a)(1-b)}{a}}, \sqrt{\frac{%
a+b-1}{a}},-\sqrt{1-b})\nonumber \\
&&\quad \mathrm{if}\quad a\not=0;\quad |B\rangle
=e^{i\beta }(0,1,0)\quad \mathrm{if}\quad a=0  \label{vectorB} \\
\beta &=&\arccos ({\frac{2\mu (A\ \mathrm{or}\ B)-\mu (A)-\mu (B)}{2\sqrt{%
(1-a)(1-b)}}}) \nonumber \\
&&\quad \mathrm{if}\quad a\not=1,b\not=1;\quad \beta \ \mathrm{%
is\ arbitrary\ if}\ a=1\ \mathrm{or}\ b=1  \label{anglebeta}
\end{eqnarray}%
We take $M({\mathbb{C}}^{3})$ the ray spanned by the vector $(0,0,1)$ in
case $\mu (A)+\mu (B)\leq 1$, and we take $M({\mathbb{C}}^{3})$ the subspace
of ${\mathbb{C}}^{3}$ spanned by vectors $(1,0,0)$ and $(0,1,0)$ in case $%
1<\mu (A)+\mu (B)$. This gives rise to a quantum-mechanical description of
the situation with probability weights $\mu (A),\mu (B)$ and $\mu (A\ 
\mathrm{or}\ B)$. Let us verify this. We have that both vectors $|A\rangle $
and $|B\rangle $ are unit vectors, since $\langle A|A\rangle =a+1-a=1$ and
either $\langle B|B\rangle ={\frac{(1-a)(1-b)}{a}}+{\frac{a+b-1}{a}}+1-b=1$
in case $a\not=0$ or $\langle B|B\rangle =1$ trivially in case $a=0.$ For
both cases of $a$, one can easily check that $\langle A|B\rangle =0$, e.g.~$%
\langle A|B\rangle =\sqrt{(1-a)(1-b)}e^{i\beta }-\sqrt{(1-a)(1-b)}e^{i\beta
}=0$ for $a\not=0$, which shows that $|A\rangle $ and $|B\rangle $ are
orthogonal. Now we only need to check whether this model yields the correct
probabilities in the expressions (\ref{quantprob}).

First, let us consider $a\neq 0, a\not=1, b\not=1.$ In case that $\mu (A)+\mu
(B)\leq 1$, we have $\langle A|M|A\rangle =1-a=\mu (A)$, $\langle
B|M|B\rangle =1-b=\mu (B)$, and $\langle A|M|B\rangle =-\sqrt{(1-a)(1-b)}%
e^{i\beta +\gamma }=\sqrt{(1-a)(1-b)}e^{i\beta }$. In case $1<\mu (A)+\mu
(B) $, we have $\langle A|M|A\rangle =a=\mu (A)$, $\langle B|M|B\rangle ={%
\frac{(1-a)(1-b)}{a}}+{\frac{a+b-1}{a}}={\frac{ab}{a}}=b=\mu (B)$, and $%
\langle A|M|B\rangle =\sqrt{a}\sqrt{{\frac{(1-a)(1-b)}{a}}}e^{i\beta }=\sqrt{%
(1-a)(1-b)}e^{i\beta }$. Hence in both cases we have $\Re \langle
A|M|B\rangle =\sqrt{(1-a)(1-b)}\cos \beta$, so that $\Re \langle A|M|B\rangle
={\frac{1}{2}}(2\mu (A\ \mathrm{or}\ B)-\mu (A)-\mu (B))$. Applying (\ref%
{anglebeta}) this gives $\mu (A\ \mathrm{or}\ B)={\frac{1}{2}}(\mu (A)+\mu
(B))+\Re \langle A|M|B\rangle $, which corresponds to (\ref{interference}). This shows that, given the values of $\mu (A)$ and $\mu (B)$, the correct value for $\mu (A\ \mathrm{or}\ B)$ is obtained in this quantum-model representation.

We note that for `extreme' values of $a,b$, this model in general does not yield correct values for the disjunction. First, let us note that the case $%
a=0$ only occurs when $\mu (A)=1$ and $\mu (B)=0$ and hence $M({\mathbb{C}}%
^{3})$ is the ray spanned by $(0,0,1)$ and $b=1$. So $|A\rangle =(0,0,1)$
and $|B\rangle =e^{i\beta }(0,1,0),$ with $\beta $ arbitrary. It is easy to check that $\langle A|M|A\rangle =1=\mu (A)$, $\langle B|M|B\rangle =0=\mu
(B)$. We have $\langle A|M|B\rangle =0$, and there is no interference. In such a case, the model is only valid if $\mu (A\ \mathrm{or}\ B)={\frac{1}{2}}%
(\mu (A)+\mu (B))={\frac{1}{2}}$. Secondly, let $a\not=0$. The case $b=1$
can also occur for $\mu (B)=1$ and $M({\mathbb{C}}^{3})$ is then the plane spanned by $(1,0,0)$ and $(0,1,0)$. It is easy to check that $\langle
A|M|A\rangle =a=\mu (A)$, $\langle B|M|B\rangle =1=\mu (B)$. Also, $\langle
A|M|B\rangle =0$, and again there is no interference. Finally, let $a=1.$
Then $|A\rangle =(1,0,0)$ and $|B\rangle =e^{i\beta }(0,\sqrt{b},-\sqrt{1-b}%
),$ with $\beta $ arbitrary. The outcome of similar calculations is again that there is no interference. In other words, as soon as $\mu (A)\in \left\{
0,1\right\} $ or $\mu (B)\in \left\{ 0,1\right\} $, the interference term in
this model vanishes and it only yields the correct numerical value if $\mu
(A\ \mathrm{or}\ B)={\frac{1}{2}}(\mu (A)+\mu (B))$. For these `extreme' cases, one needs to adopt a slightly more elaborate model, as presented in 
\cite{aerts2009}. In fact, depending on the values of $\mu (A),\mu (B)$ and $%
\mu (A\ \mathrm{or}\ B)$, the equation (\ref{anglebeta}) determines whether it is possible to construct a representation of the data in a ${\mathbb{C}}%
^{3}$ model or not. Nevertheless, for many situations encountered in `real-life' economic paradoxes like the Hawaii problem, the probabilities $%
\mu (A),\mu (B)$ and $\mu (A\ \mathrm{or}\ B)$ take `non-extreme' values such that the ${\mathbb{C}}^{3}$ model suffices.

In this ${\mathbb{C}}^{3}$ model, only $e^{i\beta }$ appears as `not a real number' in the vector $|B\rangle $. For two values of $\beta $, namely $%
\beta =0^{\circ }$ and $\beta =180^{\circ }$, $e^{i\beta }$ is a real number, which means that for these two values of $\beta $ we can make a graphical representation of the situation in ${\mathbb{R}}^{3}$. The interference effect is present for these two values of $\beta $ but can take only two values. The role of the complex numbers is to allow it to obtain any value in between these two values. In \cite{aerts2009} the vectors $%
|A\rangle $ and $|B\rangle $ and the angle $\beta $ for a number of experimental data in concept theory have been calculated. For some of the items no ${\mathbb{C}}^{3}$ model exists and one needs to extend the modeling to Fock space \cite{aerts2007b}. However, for the specific case of the Hawaii problem a quantum representation by means of the ${\mathbb{C}}^{3}$ model is possible. Indeed, we have $\mu (A)=0.54$, $\mu (B)=0.57$ and $\mu (A\ \mathrm{or}\ B)=0.32$. First, let us note that this means that this situation does not allow a classical model, since $\mu (A\ \mathrm{or}\
B)<\mu (B)$. Secondly, let us construct the ${\mathbb{C}}^{3}$ quantum model for this situation. We have $1< \mu (A)+\mu (B)=1.11$, and hence we put $a=0.54$, $b=0.57$ and $\gamma =0$. After making the calculations of equations (\ref%
{vectorA}), (\ref{vectorB}) and (\ref{anglebeta}), we obtain $|A\rangle
=(0.7348,0,0.6782)$, $|B\rangle =e^{i121.8967^{\circ
}}(0.6052,0.4513,-0.6557)$ and we take $M({\mathbb{C}}^{3})$ the subspace of ${\mathbb{C}}^{3}$ spanned by vectors $(1,0,0)$ and $(0,1,0)$. It is easy to verify that this model indeed yields the correct numerical outcomes.

\section{Concept Combinations, the disjunction effect and conceptual landscapes}

The disjunction effect has most of all been studied within the research domain of `decision theory'. Quantum game theory has been used to model it \cite{busemeyerwangtownsend2006,pothosbusemeyer2009}, and also quantum theoretical models have been worked out for it \cite{busemeyerpothosfranco2011,khrennikov2008,yukalovsornette2010} along similar lines as the models we worked out \cite{aerts2007a,aerts2007b,aerts2009,aertsaerts1994} and of which we presented a short presentation in the foregoing section. In different terms, however, the effect was studied experimentally in connection with problems occurring with the combination of concepts \cite{hampton1988}. Let us show by an example what we mean. Consider the concepts {\it Home Furnishings} and {\it Furniture} and their disjunction {\it Home Furnishings or Furniture} and the item {\it Ashtray}. Subjects estimated the membership weight of {\it Ashtray} for the concept {\it Home Furnishings} to be 0.3 and the membership weight of the item {\it Ashtray} for the concept {\it Furniture} to be 0.7. However, the membership weight of {\it Ashtray} with respect to the disjunction {\it Home Furnishings or Furniture} was estimated only 0.25, less than both weights with respect to both concepts apart \cite{hampton1988}. This deviation from a standard classical interpretation of the disjunction was called `underextension' \cite{hampton1988}. We can easily see the correspondence with the original situation of the disjunction effect, for example the Hawaii problem. What is particularly interesting, however, is that although the item {\it Ashtray} shows underextension with respect to the disjunction of the concepts {\it Home Furnishings} and {\it Furniture}, the study of concepts and their disjunction shows that on many occasions the inverse effect occurs as well. For example, with respect to the pair of concepts {\it Fruits} and {\it Vegetables} and their disjunction {\it Fruits or Vegetables}, for the item {\it Olive} the membership weights with respect to {\it Fruits}, {\it Vegetables} and {\it Fruits or Vegetables} were 0.5, 0.1 and 0.8, respectively. This `inverse disjunction effect' has been called `overextension' \cite{hampton1988}. We can prove that for these weights it is not possible to find a Kolmogorovian representation \cite{aerts2009}. This means that these weights cannot be obtained by supposing that subjects reasoned following classical logic and that the weights are the result of a lack of knowledge about the exact outcomes given by each of the individual subjects. Indeed, if 50\% of the subjects have classified the item {\it Olive} as belonging to Fruits, and 10\% have classified it as belonging to {\it Vegetables}, then following classical reasoning at most 60\% of the subjects (corresponding with the set-theoretic union of two mutually distinct sets of subjects) can classify it as belonging to `{\it Fruits or Vegetables}', while the experiment shows that $80\%$ did so. This means that these weights arise in a distinct way. Some individual subjects must necessarily have chosen {\it Olive} as a member of `{\it Fruits or Vegetables} and `not as a member' of {\it Fruits} and also `not as a member' of {\it Vegetables}, otherwise the weights 0.5, 0.1 and 0.8 would be impossible results. It is here that the previously referred to second layer of thought --- quantum-conceptual thought --- comes into play. Concretely, this means that for the item {\it Olive}, the subject, rather than reasoning in a logical way, directly wonders whether {\it Olive} is a member or not a member of `{\it Fruits or Vegetables}'. In this quantum-conceptual thought process, the subject considers `{\it Fruits or Vegetables}' as a newly emerging concept and not as a classical logical disjunction of the two concepts {\it Fruits} and {\it Vegetables} apart. As a result, the subject's choice is influenced in favor or against membership of {\it Olive} with respect to {\it Fruits or Vegetables} by the presence of the new concept `{\it Fruits or Vegetables}'. This is why we say that within the quantum-conceptual thought process it is the emergence of a new concept, i.e. the concept `{\it Fruits or Vegetables}', within the landscape of existing concepts, i.e. {\it Fruits}, {\it Vegetables} and {\it Olive}, that gives rise to the deviation from the membership weight that would be expected following classical logic. And it is the probability of deciding for or against membership that is influenced by the presence of this new concept within the landscape of existing concepts. Concretely, in this case, the subject estimates whether {\it Olive} is characteristic of the new concept `{\it Fruits or Vegetables}', evaluating whether {\it Olive} is one of those items that raise doubts as to whether it is a {\it Fruit} or a {\it Vegetable}. And since for {\it Olive} this is typically the case, its weight with respect to `{\it Fruits or Vegetables}' is relatively big, namely 0.8, compared to the weights 0.5 and 0.1 with respect to the individual concepts {\it Fruits} and {\it Vegetables}, respectively. Most importantly, it is `bigger than the sum of both' (0.8 is strictly bigger than 0.5 + 0.1), which makes a classical logical and stochastic probabilistic explanation impossible, as we explicitly proved in earlier work \cite{aerts2009}. As a consequence of this finding --- that the weights for the item {\it Olive} with respect to the pair of concepts {\it Fruits}, {\it Vegetables} and `{\it Fruits or Vegetables}' cannot be modeled within a Kolmogorovian probability structure --- there cannot exist an underlying deterministic process giving rise to these weights. This means that the conceptual thought process that takes place when a subject decides `for' or `against' membership of the item {\it Olive} with respect to the concepts {\it Fruits}, {\it Vegetables} and {\it Fruits or Vegetables} intrinsically does not follow the rules of classical logic. We have shown that the conceptual thought process resulting in the weights with respect to the different conceptual structures, when different possible decisions are considered, can be modeled by means of a quantum-mechanical probability structure \cite{aerts2009}.

Is it possible to apply the above explanation for deviations from classical logical thought, i.e. the presence of quantum-conceptual thought, to the disjunction of concepts with respect to the traditional disjunction effect, for example the Hawaii problem, so as to understand the violation of the Sure-Thing Principle? There is a set of experiments \cite{bagassimacchi2007}, although performed with a different goal from ours, which confirm that our explanation for concepts and their disjunction is also valid for the traditional disjunction effect. The experiments consider the Hawaii problem to show that the disjunction effect does not depend on the presence of uncertainty (pass or fail the exam) but on the introduction into the text-problem of a non-relevant goal \cite{bagassimacchi2007}. This indicates in a very explicit way that it is the overall conceptual landscape that gives form to the disjunction effect. More specifically, the authors point out that option $z$ (`pay a \$5 non-refundable fee in order to retain the rights to buy the vacation package at the same exceptional price the day after tomorrow after you find out whether or not you passed the exam') contains an unnecessary goal, i.e. that one needs to `pay to know', which is independent of the uncertainty condition. In this sense, their hypothesis is that the choice of option $z$ occurs as a consequence of the construction of the discourse problem itself \cite{bagassimacchi2007}. Four experiments were performed in which various modifications with respect to option $z$ were considered, ranging from (1) eliminating from the text and option $z$ any connection between the `knowledge of the outcome' and the `decision'; (2) eliminating option $z$, limiting the decision to $x$ (`buy') or $y$ (`not buy'); (3) making option $z$ more attractive; (4) render the procrastination option $z$ more onerous. The experimental results support the view that the disjunction effect does not depend on the uncertainty condition itself, but on the insertion of the misleading goal --- `paying to know' --- into the text-problem. When it is eliminated (but maintaining the uncertainty condition), the disjunction effect vanishes (exps. 1 and 2). Also, if choosing $z$ makes sense (exp. 3), most subjects choose it. If option $z$ is onerous (exp. 4), it is substantially ignored. In this sense, option $z$ is not a real alternative to $x$ and $y$, but becomes an additional premise that conveys information which changes the decisional conceptual landscape. Hence the crucial factor is the relevance of the discourse problem of which $z$ is one element, rather than certainty versus uncertainty. These results support the view that the disjunction effect appears when a suitable decisional conceptual landscape is present rather than mainly depending on the presence of uncertainty. If the specific conceptual landscape surrounding the decision situation is what lies at the origin of the disjunction effect, then this shows that what we have called `quantum-conceptual thought' is taking place during the process that gives rise to the disjunction effect. Following these experimental results, one can argue that `the specific conceptual landscape surrounding the decision situation' plays a principal role in shaping the disjunction effect.

To illustrate the above explanation directly for the case of the Hawaii problem, let us suppose that the Hawaii dilemma is not about buying or not buying a vacation package but about saying `yes' or `no' to a free Sauna and Jacuzzi relaxation weekend. In this case, the outcome may well be the opposite, with a larger than expected number of subjects saying `yes', particularly in the situation in which they do not know their exam results, because their state of anxiety may in fact make the prospect of a relaxation weekend more attractive than in the other two situations, in which they do know the results, whether pass or fail. This would be an example of the inverse disjunction effect or overextension of the disjunction. We indeed believe that in the traditional Hawaii situation, the disjunction effect, hence the underextension, is due to the fact that the idea of making a trip is appealing only if the outcome of the exam is known, whether pass or fail, whereas a free relaxation weekend is much more likely to be appealing even if the exam results have not been released yet. We have conducted ourselves experiments with the aim of detecting different pieces of the overall conceptual landscape determining the nature of the disjunction effect in the case of the Ellsberg paradox \cite{aertsdhooghesozzo2011}, illustrating the above expressed view.

\section{Conclusion}

At this stage we should point out the following. In our earlier work \cite{aerts2009,aertsdhooghe2009} we introduced the notion of `conceptual landscape'. One of the reasons for this is that our approach is grounded in a modeling of concepts and their combinations \cite{aertsbroekaertgabora2010,aertsgabora2005a,aertsgabora2005b,gaboraaerts2002}. Of course, sometimes a human decision will be made under the influence of a surrounding that cannot easily be expressed as a conceptual landscape. When we pointed out explicitly that it is `the whole' conceptual landscape which needs to be taken into account and modeled within our quantum modeling scheme, we could have expressed this in a more complete way by using the notion of `worldview' as understood and analyzed in detail in the Leo Apostel Center \cite{aertsaposteldemoorhellemansmaexvanbellevanderveken1994,aertsaposteldemoorhellemansmaexvanbellevanderveken1995,aertsvanbellevanderveken1999}. These are all possible elements of a worldview surrounding a given situation that can influence a human decision being made in this situation. Of course, if the elements of the surrounding worldview considered can be expressed conceptually, i.e. if they are of the form of a conceptual landscape, then these elements can be taken into account by means of the quantum modeling scheme we have developed in earlier work for concepts and their combinations \cite{aerts2007a,aerts2007b,aerts2009,aertsdhooghe2009,aertsgabora2005a,aertsgabora2005b,gaboraaerts2002}. This being the case, we are already able to grasp an enormously important and also substantial part of the dynamics generated by the totality of the worldview influence, which is the reason why we have focused on this in our previous work.

\end{document}